\documentclass[12 pt, amsfonts,amsmath, amssymb,color]{article}
\usepackage{authblk}
\def\baselinestretch{1.0}
\usepackage{amsmath,amssymb,amsfonts,latexsym,float,graphics,epsfig}
\usepackage{subfig}
\usepackage{verbatim}
\usepackage{relsize}
\usepackage{hyperref}
\usepackage{graphicx}
\usepackage{textcomp}
\usepackage{bm}
\usepackage{epstopdf}
\usepackage{slashed}
\usepackage{color}
\usepackage{tikz-cd}
\usepackage[utf8]{inputenc}
\usepackage[T1]{fontenc}

\def\be{\begin{equation}}
\def\ee{\end{equation}}
\def\bea{\begin{eqnarray}}
\def\eea{\end{eqnarray}}

\begin{document}

\renewcommand\theequation{\arabic{section}.\arabic{equation}}
\catcode`@=11 \@addtoreset{equation}{section}
\newtheorem{axiom}{Definition}[section]
\newtheorem{theorem}{Theorem}[section]
\newtheorem{axiom2}{Example}[section]
\newtheorem{lem}{Lemma}[section]
\newtheorem{prop}{Proposition}[section]
\newtheorem{cor}{Corollary}[section]
\newtheorem{res}{Results}[section]

\newcommand{\ben}{\begin{equation*}}
\newcommand{\een}{\end{equation*}}

\let\endtitlepage\relax

\begin{titlepage}
\begin{center}
\renewcommand{\baselinestretch}{1.5}  

\vspace*{-0.5cm}

{\Large {On novel Hamiltonian descriptions of some}}\\
 {\Large {three-dimensional non-conservative systems}}
 
\vspace{5mm}
\renewcommand{\baselinestretch}{1}  

\centerline{{\bf Aritra Ghosh$^*$\footnote{aritraghosh500@gmail.com, ag34@iitbbs.ac.in}, Anindya Ghose-Choudhury$^\dagger$\footnote{aghosechoudhury@gmail.com}, Partha Guha$^\ddagger$\footnote{partha.guha@ku.ac.ae}}}

\vspace{7mm}
\normalsize
\text{$^*$School of Basic Sciences, Indian Institute of Technology Bhubaneswar,}\\
\text{Jatni, Khurda, Odisha 752050, India}\\

\vspace{1.5mm}
\text{$^\dagger$Department of Physics, Diamond Harbour Women’s University,}\\
\text{D.H. Road, Sarisha, West Bengal 743368, India}\\

\vspace{1.5mm}
\text{$^\ddagger$Department of Mathematics, Khalifa University of Science and Technology,}\\
\text{Main Campus, P.O. Box -127788, Abu Dhabi, United Arab Emirates}\\

\vspace{7mm}

\begin{abstract}
We present novel Hamiltonian descriptions of some three-dimensional systems including two well-known systems describing the three-wave-interaction problem and some well-known chaotic systems, namely, the Chen, L\"u, and Qi systems. We show that all of these systems can be described in a Hamiltonian framework in which the Poisson matrix $\mathcal{J}$ is supplemented by a resistance matrix $\mathcal{R}$. While such resistive-Hamiltonian systems are manifestly non-conservative, we construct higher-degree Poisson matrices via the Jordan product as $\mathcal{N} = \mathcal{J} \mathcal{R} + \mathcal{R} \mathcal{J}$, thereby leading to new bi-Hamiltonian systems. Finally, we discuss conformal Hamiltonian dynamics on Poisson manifolds and demonstrate that by appropriately choosing the underlying parameters, the reduced three-wave-interaction model as well as the Chen and L\"u systems can be described in this manner where the concomitant non-conservative part of the dynamics is described with the aid of the Euler vector field. 
\end{abstract}
\end{center}
\vspace*{0cm}


\end{titlepage}

\section{Introduction}
The framework of Hamiltonian dynamics is a powerful tool describing the time evolution of certain dynamical systems using `energy-based' principles. This approach is capable of providing deep insights into the structure and symmetries of the laws of the physical world and also provides the grounds for the elegant geometrical formulations of mechanics. In classical mechanics \cite{arnold}, Hamiltonian dynamics is formulated on the phase space (which is of even dimension) on which the local coordinates are the so-called generalized coordinates and their conjugate momenta; geometrically, such a phase space can be understood to be the cotangent bundle $T^*\mathcal{Q}$ of the configuration space $\mathcal{Q}$; the former naturally admits a symplectic structure that allows one to define Poisson brackets between smooth functions on the phase space \cite{arnold}. Such Poisson brackets (also called canonical Poisson brackets) have important implications in physics and are often one of the primary ingredients of quantization. On the other hand, it is straightforward to generalize the Hamiltonian framework to general phase spaces which do not necessarily have the symplectic structure but can still be endowed with a Poisson bracket \cite{poisson1}. Such brackets (also called non-canonical Poisson brackets) have intriguing implications and allow for a Hamiltonian description of certain systems in fluid dynamics \cite{morr,fluid1,fluid2}, plasma physics \cite{morr1,morr2}, statistical physics \cite{stat}, elastic structures \cite{elas1,elas2}, etc., just to name a few. The underlying idea is to consider the phase space to be a Poisson manifold which may also be closely related to Nambu-Poisson geometry \cite{nambu0,nambu,nambu2}. \\

In the recent decades, there has been an interest in understanding the mathematical structure of three-dimensional dynamical systems \cite{3w2,3w1,chen,lu1,lu2,qi,t}, most notably in context of the theory of chaos \cite{chen,lu1,lu2,qi,t}. The Hamiltonian framework has played a key role in this \cite{gumralimp,gumral,gao,ogul,ogul1,ogul2,ogul3}. In particular, it was demonstrated by G\"umral and Nutku \cite{gumralimp} (see also, \cite{gumral}) that every Hamiltonian system in three dimensions is mutually bi-Hamiltonian and they derived non-canonical bi-Hamiltonian structures of the Lorenz system for certain values of its parameters. Moreover, Gao \cite{gao} showed that for a three-dimensional dynamical system with one time-independent first integral to admit a Hamiltonian description, the necessary and sufficient condition is the existence of a Jacobi last multiplier \cite{JLM,JLM1} which naturally indicates the integrability of the system (at least locally); the framework was then utilized to present a Hamiltonian formulation of the Lotka-Volterra system. Several other well-known three-dimensional systems admitting chaotic behavior, namely, the Chen system \cite{chen}, L\"u (and modified L\"u) system \cite{lu1,lu2}, Qi system \cite{qi}, and T system \cite{t} can be associated with bi-Hamiltonian character \cite{ogul}, including a possible formulation using Nambu-Poisson structures \cite{ogul1}. Other three-dimensional systems, including the reduced three-wave-interaction model and the Rabinovich system \cite{3w2,3w1} have been associated with Hamiltonian formulations using the so-called Nambu-metriplectic structures \cite{ogul2}. 

\subsection{Motivation, results, and outline}
Motivated by the above-mentioned developments including the large number of interesting examples available in the literature, we attempt to put forward a Hamiltonian framework suiting several of these by introducing mathematical structures inspired from the framework of port-Hamiltonian systems \cite{port,port1}, albeit without source terms (also known as external ports). In particular, we will show that by introducing a so-called resistance matrix, the Chen, L\"u (including a modified version), and Qi systems can all be described by such a Hamiltonian structure. This construction is also shown to work for the reduced three-wave interaction model and the Rabinovich system. Furthermore, we will describe higher-degree Poisson matrices leading naturally to bi-Hamiltonian structures and point out certain similarities with the Jordan product and the Jordan identity \cite{jacobson,jordan}. Finally, we shall describe conformal Hamiltonian dynamics \cite{conf1,conf2} (see also, \cite{ogul1}) on Poisson manifolds and demonstrate that by choosing the underlying parameters appropriately, some of the above-mentioned examples, namely, the reduced three-wave-interaction model as well as the Chen and L\"u systems can be described by this framework.\\

In particular, our results can be summarized as follows: 

\begin{enumerate}
\item The well-known three-dimensional systems, namely, the reduced three-wave interaction model, the Rabinovich system, the Chen, L\"u, and Qi systems can all be described in a Hamiltonian framework in which the Poisson part of the dynamics is supplemented by a resistive part that describes the non-conservative part of the dynamics. 

\item Beginning with a Poisson matrix $\mathcal{J}$ and a suitable resistance matrix $\mathcal{R}$, we shall construct new Poisson matrices via the Jordan-like product $\mathcal{J} \mathcal{R} + \mathcal{R} \mathcal{J}$. This allows one to derive novel (conservative) bi-Hamiltonian systems starting with certain known non-conservative systems. Transformations inspired from Jordan homomorphisms allow one to derive further new systems. 

\item Some of the well-known three-dimensional systems, namely, the reduced three-wave interaction model as well as the Chen and L\"u systems are shown to be described by conformal Hamiltonian dynamics on Poisson manifolds for appropriate choices of the underlying parameters. 
\end{enumerate}

With this introduction, let us present the organization of this paper. The preliminaries of Poisson structures on smooth manifolds will be reviewed in the following section [Sec. (\ref{revsec})] while Hamiltonian systems with resistance matrices will be introduced in Sec. (\ref{ressec}) in which several three-dimensional examples will also be worked out. This is followed by a description of higher-degree Poisson matrices and the associated bi-Hamiltonian structures in Sec. (\ref{highersec}). Then, in Sec. (\ref{confsec}), we will discuss the framework of conformal Hamiltonian dynamics on Poisson manifolds and apply it to some three-dimensional examples. Finally, we will conclude the paper in Sec. (\ref{dissec}).

\section{Hamiltonian dynamics on Poisson manifolds: A review}\label{revsec}
A Poisson manifold is a smooth manifold $\mathcal{M}$ endowed with a Poisson structure \cite{poisson1}, i.e., the pair $(\mathcal{M},\Lambda)$, where $\Lambda$ is a bi-vector field which satisfies the following integrability condition:
\begin{equation}\label{int}
[\Lambda, \Lambda] = 0, 
\end{equation} with $[\cdot,\cdot]$ being the Schouten-Nijenhuis bracket \cite{poisson1}. For two functions $f,g \in C^\infty(\mathcal{M},\mathbb{R})$,  the Poisson bi-vector allows one to define a local Lie bracket on $C^\infty(\mathcal{M},\mathbb{R})$ as $\Lambda(df, dg) = \{f,g\}$ called the Poisson bracket which is obviously skew-symmetric. Additionally, the condition (\ref{int}) enforces the so-called Jacobi identity $\{\{f,g\},h\} + \{\{h,f\},g\} + \{\{g,h\},f\} = 0$, $\forall f,g,h \in C^\infty(\mathcal{M},\mathbb{R})$. \\

The definition $\Lambda(df, dg) = \{f,g\}$ suggests that the Poisson bracket acts as a derivation, i.e., it satisfies the Liebnitz identity; it can therefore define a vector field. Given a Poisson manifold $(\mathcal{M},\Lambda)$ and a function $H \in C^\infty(\mathcal{M}, \mathbb{R})$, the Hamiltonian vector field (corresponding to $H$) is defined as
\begin{equation}\label{Hamdef}
X_H = \{\cdot, H\}. 
\end{equation} We will then call $H$ the Hamiltonian. In local coordinates, one can write
\begin{equation}
\Lambda = \mathcal{J}_{ij} \frac{\partial}{\partial x_i} \wedge \frac{\partial}{\partial x_j},
\end{equation} where $i,j \in \{1,2,\cdots, n\}$ with ${\rm dim} (\mathcal{M}) = n$. The $n \times n$ matrix $\mathcal{J}$ is called the Poisson matrix which is skew-symmetric and must also be compatible with the integrability condition (\ref{int}); this requires
\begin{equation}\label{jac0}
\mathcal{J}^{i[j} \partial_i \mathcal{J}^{kl]} = 0,
\end{equation} where $i,j,k,l \in \{1,2,\cdots,n\}$ and $\partial_i = \partial/\partial x_i$. The equations of motion, i.e., $\dot{x}_i = X_H (x_i)$ can be expressed in the following manner using matrices and column vectors \cite{gumralimp}:
\begin{equation}\label{Hamdef2}
\dot{\mathbf{x}} = \mathcal{J} (\nabla H),
\end{equation} where $\mathbf{x} = (x_1 \quad x_2 \cdots x_n)^T$ and $\nabla$ is the usual del operator. An immediate consequence of (\ref{Hamdef}) or (\ref{Hamdef2}) is that $\frac{dH}{dt} = X_H(H) =  0$, i.e., the Hamiltonian $H$ is conserved. Thus, Hamiltonian dynamics on Poisson manifolds describes conservative systems that conserve the Hamiltonian function; notice that symplectic manifolds encountered in classical mechanics are all endowed with a Poisson structure due to the symplectic two-form \cite{arnold}. It is then naturally implied that the description of non-conservative systems is possible only upon introducing additional elements into the formalism.

\subsection{Three-dimensional systems}\label{3drev}
When considering three-dimensional dynamical systems, i.e., where the phase space is of (real) dimension three, the Jacobi identity takes an intriguing form. It should be remarked that the space of $3 \times 3$ skew-symmetric matrices is isomorphic to the space of three-dimensional vectors as \cite{gumralimp}
\begin{equation}\label{isomorphism}
\mathcal{J} = \begin{pmatrix}
0 & -J_z & J_y \\
J_z & 0 & -J_x \\
-J_y & J_x & 0
\end{pmatrix} \leftrightarrow \mathbf{J} = (J_x,J_y,J_z),
\end{equation} where $\mathbf{J}$ should be distinguished from $\mathcal{J}$. If we demand that $\mathcal{J}$ is a Poisson matrix on $\mathbb{R}^3$, then the integrability condition (\ref{int}) leading to the Jacobi identity takes the following form (see for example, \cite{gumral}):
\begin{equation}\label{jacobi}
\mathbf{J} \cdot (\nabla \times \mathbf{J}) = 0,
\end{equation} and the dynamical equations represented by equation (\ref{Hamdef2}) are given by
\begin{equation}\label{Hamdef3}
\dot{\mathbf{x}} = \mathbf{J} \times \nabla H. 
\end{equation}
Notice that the general solution of the equation (\ref{jacobi}) is locally of the form (see \cite{gumralimp} and references therein)
\begin{equation}\label{JLM}
\mathbf{J} = \frac{1}{M} (\nabla \overline{H}), 
\end{equation} for some function $\overline{H} \in C^\infty(\mathcal{M},\mathbb{R})$ and $M$ is an integrating factor called the Jacobi last multiplier \cite{JLM,JLM1}. Thus, a Poisson system may admit a non-trivial divergence\footnote{For an arbitrary vector field $X$ on a smooth and orientable manifold $\mathcal{M}$ with volume-form $\Omega = dx_1 \wedge dx_2 \wedge \cdots \wedge dx_n$, the divergence can be defined as $\pounds_{X} \Omega = {\rm div}(X) \Omega$ (see \cite{AG} and references therein). In the notation of three-dimensional vectors in $\mathbb{R}^3$ as employed in this paper, this agrees with that obtained by $\nabla \cdot \dot{\mathbf{x}}$.} unlike a Hamiltonian system on a symplectic manifold \cite{arnold} because
\begin{equation}\label{Hamdiv3d}
(\nabla M)\cdot \dot{\mathbf{x}} + M (\nabla \cdot \dot{\mathbf{x}}) = 0,
\end{equation} i.e., if $M$ is not a constant, then $ \nabla \cdot \dot{\mathbf{x}} \neq 0$ and vice versa; notice that $M$ is a constant if and only if $\nabla \times \mathbf{J} = 0$. The above discussion reveals that we can re-write the vector equation (\ref{Hamdef3}) using (\ref{JLM}) such that
\begin{equation}\label{HbiHamiltonian}
M \dot{\mathbf{x}} = (\nabla \overline{H}) \times (\nabla H),
\end{equation} and which indicates a bi-Hamiltonian character. In fact, defining a second Poisson vector as
\begin{equation}
\overline{\mathbf{J}} = \frac{1}{M}  (\nabla H), 
\end{equation} one can reproduce the dynamics (\ref{Hamdef3}) or (\ref{HbiHamiltonian}) if the following compatibility condition is obeyed \cite{gumral,ogul1}:
\begin{equation}\label{comp}
\mathbf{J}\cdot(\nabla \times \overline{\mathbf{J}}) = \overline{\mathbf{J}}\cdot(\nabla \times \mathbf{J}) = 0. 
\end{equation}
It is easy to see that under the resulting dynamics, both $H$ and $\overline{H}$ are conserved. The dynamics (\ref{HbiHamiltonian}) can be represented in the Nambu-Poisson framework \cite{nambu0,nambu,nambu2} by recognizing that the dynamics of an arbitrary function $F \in C^\infty(\mathcal{M}, \mathbb{R})$ is given by (see also, \cite{gumralimp,gumral,ogul,ogul1,ogul2,ogul3})
\begin{equation}
\dot{F} = \frac{1}{M} (\nabla F) \cdot (\nabla \overline{H}) \times (\nabla H) \equiv \{F,\overline{H},H\},
\end{equation}
where $\{\cdot, \cdot, \cdot\}$ is the Nambu three-bracket which satisfies a generalized and stronger version of the Jacobi identity called the fundamental identity which was introduced in \cite{nambu}. Note that any three-dimensional bi-Hamiltonian system is also a Nambu-Poisson system.

\section{Resistive and port-Hamiltonian systems}\label{ressec}
In this section, we will introduce a modified class of Hamiltonian systems which we will call resistive-Hamiltonian systems taking inspiration from the notion of resistance in electrical circuits. To this end, consider an $LC$-circuit, i.e., a circuit with no sources but with an inductor $L$ and a capacitor $C$ connected in series. Denoting the charge on the capacitor to be $Q$, one gets the following differential equations from Kirchhoff's voltage rule:
\begin{equation}\label{kir1}
 \dot{Q} = I, \quad \quad L \dot{I} = - \frac{Q}{C} . 
\end{equation}
Defining $\mathbf{x} = (Q \quad I)^T$ on a two-dimensional Poisson manifold and $H = \frac{I^2}{2} + \frac{Q^2}{2 LC}$, equations (\ref{kir1}) can be cast in the form of (\ref{Hamdef2}) if we choose
\begin{equation}
\mathcal{J} = \begin{pmatrix}
0 & 1 \\
-1 & 0 
\end{pmatrix}. 
\end{equation} Thus, the conservative dynamics of $LC$-oscillations can be described by a Hamiltonian as is well known. Let us now introduce a resistive element with resistance $R$ into the circuit. The Kirchhoff's voltage rule shall then give
\begin{equation}\label{kir2}
 \dot{Q} = I, \quad \quad L \dot{I} = - R I - \frac{Q}{C},
\end{equation} where we have used Ohm's law to write the voltage drop across the resistance $R$ as $-I R$. The system (\ref{kir2}) can no longer be expressed in the form of (\ref{Hamdef2}) for a skew-symmetric matrix $\mathcal{J}$; however, one can write down a modified evolution equation for the same Hamiltonian $H = \frac{I^2}{2} + \frac{Q^2}{2 LC}$ as
\begin{equation}
\dot{\mathbf{x}} = (\mathcal{J} - \mathcal{R}) (\nabla H), \quad \quad \mathcal{J} = \begin{pmatrix}
0 & 1 \\
-1 & 0 
\end{pmatrix}, \quad \quad 
\mathcal{R} = \begin{pmatrix}
0 & 0 \\
0 & \frac{R}{L} 
\end{pmatrix}. 
\end{equation}
Moreover, if one now includes a source term, say, a voltage source $V$, (\ref{kir2}) is modified to
\begin{equation}\label{kir22}
 \dot{Q} = I, \quad \quad L \dot{I} = - R I - \frac{Q}{C} + V,
\end{equation} and which can be represented in the matrix form as
\begin{equation}
\dot{\mathbf{x}} = (\mathcal{J} - \mathcal{R}) (\nabla H) + \mathcal{V}, \quad \quad \mathcal{J} = \begin{pmatrix}
0 & 1 \\
-1 & 0 
\end{pmatrix}, \quad \quad 
\mathcal{R} = \begin{pmatrix}
0 & 0 \\
0 & \frac{R}{L} 
\end{pmatrix}, \quad \quad \mathcal{V} = V \begin{pmatrix}
0 \\
1   
\end{pmatrix}.
\end{equation}
This brings us to port-Hamiltonian systems \cite{port,port1}:
\begin{axiom}
Consider a Poisson manifold $(\mathcal{M},\Lambda)$ with a Poisson matrix $\mathcal{J}$. Given a Hamiltonian $H \in C^\infty(\mathcal{M}, \mathbb{R})$ and a symmetric matrix $\mathcal{R}$, a port-Hamiltonian system takes the following generic form:
\begin{equation}\label{portHam}
\dot{\mathbf{x}} = (\mathcal{J} - \mathcal{R}) (\nabla H) + \mathcal{V},
\end{equation} where $\mathbf{x} = (x_1 \quad x_2 \cdots x_n)^T$ with $x_1, x_2, \cdots, x_n$ being local coordinates on $\mathcal{M}$ and $\mathcal{V}$ is a suitable column vector representing sources, feedbacks, etc, containing the effect of the so-called external ports. 
\end{axiom}

It must be mentioned that in the theory of port-Hamiltonian systems, the skew-symmetric matrix $\mathcal{J}$ is not required to satisfy the Jacobi identity (\ref{jac0}) although in many cases it does \cite{port}; for our purposes, we will consider $\mathcal{J}$ to be a Poisson matrix. Port-Hamiltonian systems are often associated with certain algebraic conditions which give rise to the almost Dirac structures.  A special case of the above class of systems are the ones with no external ports, i.e, $\mathcal{V} = (0 \quad 0 \cdots 0)^T$. These are then associated with losses (or gains) described by the resistance matrix alone and will be dubbed `resistive-Hamiltonian systems'. They can be defined formally as follows:

\begin{axiom}
Consider a Poisson manifold $(\mathcal{M},\Lambda)$ with a Poisson matrix $\mathcal{J}$. Given a Hamiltonian $H \in C^\infty(\mathcal{M}, \mathbb{R})$ and a symmetric matrix $\mathcal{R}$, a resistive-Hamiltonian system is defined as
\begin{equation}\label{resHam}
\dot{\mathbf{x}} = (\mathcal{J} - \mathcal{R}) (\nabla H),
\end{equation} where $\mathbf{x} = (x_1 \quad x_2 \cdots x_n)^T$ with $x_1, x_2, \cdots, x_n$ being local coordinates on $\mathcal{M}$. 
\end{axiom}
When described by a vector field, we will denote it with $\mathcal{X}_H$ as opposed to the conservative Hamiltonian vector field denoted by $X_H$. It should be noted that the matrix $\mathcal{R}$, being symmetric, can describe a symmetric bracket $(f,g) = ( \nabla f)^T \mathcal{R} (\nabla g)$, where $f,g \in C^\infty(\mathcal{M}, \mathbb{R})$. Such a bracket clearly does not satisfy the Jacobi identity but does satisfy bilinearity and the Leibniz identity, both due to the appearance of the differential operator $\nabla$. It is therefore a Leibniz bracket (see for example, \cite{morr,morr1,morr2,guha_metri}). In what follows, we will show that several well-known three-dimensional systems fall into the class of resistive-Hamiltonian systems defined above.  

\subsection{Three-wave-interaction models}
We will consider the two three-wave-interaction models studied in \cite{3w2,3w1}, both of which can be cast into the form of a resistive-Hamiltonian system. 

\subsubsection{Reduced three-wave-interaction model} 
The reduced three-wave-interaction model arises in physical situations where three quasi-synchronous waves interact in a plasma with quadratic nonlinearities. The system is described by the following set of first-order equations: 
\begin{equation}\label{red3w}
\dot{x} = -2y^2 + \gamma x + z + \delta y, \quad \quad \dot{y} = 2xy + \gamma y - \delta x, \quad \quad \dot{z} = - 2xz - 2z,
\end{equation} where $\gamma$ and $\delta$ are real constants. The equations of motion can be expressed directly in the form of equation (\ref{resHam}) by identifying
\begin{equation}
H = x^2 + y^2 + z, \quad \quad \mathcal{J} = \begin{pmatrix}
0 &  -y + \frac{\delta}{2} &  z \\
y - \frac{\delta}{2} &  0 &  0 \\
-z & 0 & 0 \\
\end{pmatrix}, \quad \quad 
\mathcal{R} =  \begin{pmatrix}
-\frac{\gamma}{2} &  0 &  0 \\
0 &  -\frac{\gamma}{2} & 0 \\
0  & 0 & 2 z \\
\end{pmatrix}.
\end{equation}
It is straightforward to verify the Jacobi identity (\ref{jacobi}). Moreover, one finds that the Hamiltonian $H$ is not conserved as its evolution is given by
\begin{equation}
\mathcal{X}_H(H) = 2\gamma(x^2 + y^2) - 2z,
\end{equation} and the divergence of the vector field $\mathcal{X}_H$ is
\begin{equation}
{\rm div}(\mathcal{X}_H) = 2(\gamma - 1),
\end{equation}
the last result indicating that the dynamics does not preserve the volume-form $\Omega = dx \wedge dy \wedge dz$ in $\mathbb{R}^3$. One can check that the corresponding Poisson vector is irrotational, i.e., $\nabla \times \mathbf{J} = 0$, meaning that the divergence above is purely due to the resistive part of the time evolution as described by the resistance matrix $\mathcal{R}$. 

\subsubsection{Rabinovich system} The Rabinovich system is described by the set of equations
\begin{equation}\label{rab}
\dot{x} = qy - k_1 x + yz, \quad \quad \dot{y} = qx - k_2 y - xz, \quad \quad \dot{z} = - k_3 z + xy,
\end{equation} where the real constants $k_1$, $k_2$, and $k_3$ are the damping rates while $q$ is a real constant proportional to the driving amplitude of the feeder wave \cite{3w2}. A straightforward calculation reveals that this system can be cast in the form of (\ref{resHam}) if we identify
\begin{equation}
H = \frac{x^2+ y^2 +z^2}{2}, \quad \quad \mathcal{J} = \begin{pmatrix}
0 &  0 & 0 \\
0 &  0 &  -\frac{x}{2} \\
0 & \frac{x}{2} & 0 \\
\end{pmatrix}, \quad \quad 
\mathcal{R} =  \begin{pmatrix}
k_1 &  -q &  -y \\
-q &  k_2 & \frac{x}{2} \\
-y  & \frac{x}{2} & k_3 \\
\end{pmatrix}.
\end{equation}
Notice that the Poisson structure satisfies the Jacobi identity (\ref{jacobi}). In this case, the evolution of the Hamiltonian is given by
\begin{equation}
\mathcal{X}_H(H) = - k_1 x^2 - k_2 y^2 - k_3z^2 + 2q xy + xyz,
\end{equation} while the divergence is also non-trivial and reads
\begin{equation}
 {\rm div} (\mathcal{X}_H) = -(k_1 + k_2 + k_3),
\end{equation} meaning that the volume-form $\Omega = dx \wedge dy \wedge dz$ in $\mathbb{R}^3$ is non-conserved. It is easy to check that the corresponding Poisson vector is irrotational, i.e., $\nabla \times \mathbf{J} = 0$, implying that the divergence above is purely due to the resistive part of the time evolution as described by the resistance matrix $\mathcal{R}$.

\subsection{Chaotic systems}
We will now take the chaotic systems, namely, the Chen, L\"u, modified L\"u, and Qi systems \cite{chen,lu1,lu2,qi} and show that all of them can be cast in the form of resistive-Hamiltonian systems. In $\mathbb{R}^3$, all of them can be derived from the Hamiltonian ($\alpha$ is a real constant)
\begin{equation}\label{Hamchaos}
H = \frac{x^2}{2} - \alpha z,
\end{equation} for different choices of $\mathcal{J}$ and $\mathcal{R}$. Let us begin with the Chen system.  

\subsubsection{Chen system}
The Chen system \cite{chen} is given by the set of equations
\begin{equation}\label{chen}
\dot{x} = \alpha y - \alpha x, \quad \quad \dot{y} = (\gamma-\alpha)x + \gamma y - xz, \quad \quad \dot{z} = xy - \beta z,
\end{equation} where $\alpha$, $\beta$, and $\gamma$ are real constants. The above system can be cast in the form of (\ref{resHam}) for the Hamiltonian (\ref{Hamchaos}) with
\begin{equation}
\mathcal{J} = \begin{pmatrix}
0 &  z &  -y \\
-z &  0 &  0 \\
y  & 0 & 0 \\
\end{pmatrix}, \quad \quad 
\mathcal{R} = \frac{1}{\alpha} \begin{pmatrix}
\alpha^2 &  \alpha(\alpha - \gamma) &  0 \\
\alpha(\alpha - \gamma) &  0 &  \gamma y \\
0  & \gamma y & -\beta z \\
\end{pmatrix}.
\end{equation} Here, one has 
\begin{equation}
\mathcal{X}_H(H) = - \alpha (x^2 - \beta z), \quad \quad {\rm div}(\mathcal{X}_H) = -\alpha - \beta + \gamma. 
\end{equation}
One can verify that $\nabla \times \mathbf{J} = 0$, meaning that the Jacobi identity (\ref{jacobi}) is automatically satisfied but this moreover indicates that the non-trivial divergence of the dynamical vector field $\mathcal{X}_H$ is entirely due to the resistive part of the time evolution as described by the resistance matrix $\mathcal{R}$. 

\subsubsection{L\"u system}
The L\"u system is described as \cite{lu1,lu2}
\begin{equation}\label{lu}
\dot{x} = \alpha y - \alpha x, \quad \quad \dot{y} = \gamma y - xz, \quad \quad \dot{z} = xy - \beta z,
\end{equation} where $\alpha$, $\beta$, and $\gamma$ are real constants. It can be expressed in the form of (\ref{resHam}) for the Hamiltonian (\ref{Hamchaos}) with
\begin{equation}
\mathcal{J} = \begin{pmatrix}
0 &  z &  -y \\
-z &  0 &  0 \\
y  & 0 & 0 \\
\end{pmatrix}, \quad \quad 
\mathcal{R} = \frac{1}{\alpha} \begin{pmatrix}
\alpha^2 &  0 &  0 \\
0 &  0 &  \gamma y \\
0  & \gamma y & -\beta z \\
\end{pmatrix}.
\end{equation} 
Notice that the form of the Poisson matrix is the same as that taken for the Chen system. For this example, one can verify that 
\begin{equation}
\mathcal{X}_H(H) = - \alpha (x^2 - \beta z), \quad \quad {\rm div}(\mathcal{X}_H) = -\alpha - \beta + \gamma. 
\end{equation}
It may be checked that $\nabla \times \mathbf{J} = 0$, implying that the Jacobi identity (\ref{jacobi}) is automatically satisfied but this moreover indicates that the non-trivial divergence of the dynamical vector field $\mathcal{X}_H$ is entirely due to the resistive part of the time evolution as described by the resistance matrix $\mathcal{R}$. 

\subsubsection{Modified L\"u system}
A modified variant of the L\"u system is given by
\begin{equation}
\dot{x} = \alpha y - \alpha x + yz, \quad \quad \dot{y} = \gamma y - xz, \quad \quad \dot{z} = xy - \beta z,
\end{equation}
where $\alpha$, $\beta$, and $\gamma$ are real constants. This system can be expressed in the form (\ref{resHam}) for the Hamiltonian (\ref{Hamchaos}) with
\begin{equation}
\mathcal{J} = \begin{pmatrix}
0 &  z &  -y\left( 1 + \frac{z}{2\alpha}\right) \\
-z &  0 &  0 \\
y\left(1 + \frac{z}{2\alpha}\right)  & 0 & 0 \\
\end{pmatrix}, \quad \quad 
\mathcal{R} = \frac{1}{\alpha} \begin{pmatrix}
\alpha^2 &  0 &  \frac{yz}{2} \\
0 &  0 &  \gamma y \\
\frac{yz}{2}  & \gamma y & -\beta z \\
\end{pmatrix}.
\end{equation} 
For this example, one can verify that
\begin{equation}
\mathcal{X}_H(H) = - \alpha (x^2 - \beta z) + xyz, \quad \quad {\rm div}(\mathcal{X}_H) = -\alpha - \beta + \gamma. 
\end{equation}
In this case, one has $\nabla \times \mathbf{J} \neq 0$ but $\mathbf{J}\cdot(\nabla \times \mathbf{J}) = 0$, meaning that the Jacobi identity (\ref{jacobi}) is satisfied.  

\subsubsection{Qi system}
The Qi system \cite{qi} is given by 
\begin{equation}
\dot{x} = \alpha y - \alpha x + yz, \quad \quad \dot{y} = \gamma x - xz - y, \quad \quad \dot{z} = xy - \beta z,
\end{equation}
where $\alpha$, $\beta$, and $\gamma$ are real constants. It can be expressed in the form (\ref{resHam}) for the Hamiltonian (\ref{Hamchaos}) with
\begin{equation}
\mathcal{J} = \begin{pmatrix}
0 &  z &  -y \left(1 + \frac{z}{2\alpha}\right) \\
-z &  0 &  0 \\
y \left(1 + \frac{z}{2\alpha}\right)  & 0 & 0 \\
\end{pmatrix}, \quad \quad 
\mathcal{R} = \frac{1}{\alpha} \begin{pmatrix}
\alpha^2 &  -\alpha \gamma &  \frac{yz}{2} \\
-\alpha \gamma &  0 &  -y \\
 \frac{yz}{2}  & -y & -\beta z \\
\end{pmatrix}.
\end{equation} 
Notice that the form of the Poisson matrix is the same as that taken for the modified L\"u system. One can verify for this system that
\begin{equation}
\mathcal{X}_H(H) = - \alpha (x^2 - \beta z) + xyz, \quad \quad {\rm div}(\mathcal{X}_H) = -(\alpha + \beta + 1). 
\end{equation}
In this case one has $\nabla \times \mathbf{J} \neq 0$ but $\mathbf{J}\cdot(\nabla \times \mathbf{J}) = 0$, meaning that the Jacobi identity (\ref{jacobi}) is satisfied.

\section{Higher-degree Poisson matrices and Jordan products}\label{highersec}
In this section, we will show how the Poisson matrix $\mathcal{J}$ and the resistance matrix $\mathcal{R}$ of a resistive-Hamiltonian system can be combined to produce a bi-Hamiltonian system in three dimensions. Starting with a $\mathcal{J}$ and $\mathcal{R}$, one can define
\begin{equation}
\mathcal{N} = \mathcal{J} \mathcal{R} + \mathcal{R} \mathcal{J},
\end{equation} and which is manifestly skew-symmetric. Such a combination can be motivated from the point of view of the Jordan product between two matrices $a$ and $b$ as $a \circ b \equiv \frac{ab + ba}{2}$ \cite{jacobson,jordan}. Taking 
\begin{equation}\label{genJR}
\mathcal{J} = \begin{pmatrix}
0 & -J_z & J_y \\
J_z & 0 & -J_x \\
-J_y & J_x & 0
\end{pmatrix}, \quad \quad \mathcal{R} = \begin{pmatrix}
R_1 & A & B \\
A & R_2 & C \\
B & C & R_3
\end{pmatrix},
\end{equation} a direct matrix multiplication gives
\begin{equation}
\mathcal{N} = \begin{pmatrix}
0 & -N_z & N_y \\
N_z & 0 & -N_x \\
-N_y & N_x & 0
\end{pmatrix},
\end{equation}
where
\begin{eqnarray}
N_x &=& J_x(R_2 + R_3) - J_y A - J_z B, \label{Ncomp} \\
N_y &=& -J_x A + J_y(R_1 + R_3) - J_z C, \nonumber \\
N_z &=& -J_x B - J_y C + J_z(R_1 + R_2). \nonumber
\end{eqnarray} 
Under the isomorphism (\ref{isomorphism}), one can associate a three-dimensional Poisson vector $\mathbf{N} = (N_x,N_y,N_z)$ and then the requirement (\ref{jacobi}) that $\mathbf{N} \cdot(\nabla \times \mathbf{N}) = 0$ (Jacobi identity) implies additional restrictions on $\mathcal{J}$ and $\mathcal{R}$. We have the following result:

\begin{prop}
For the examples considered in this paper, the Jacobi identity is satisfied by $\mathcal{N}=  \mathcal{J} \mathcal{R} + \mathcal{R} \mathcal{J}$ for
\begin{enumerate}
\item \textbf{Reduced three-wave-interaction model:} arbitrary real values of the parameters.
\item \textbf{Rabinovich system:} $k_2 = - k_3$ with arbitrary real values of the other parameters. 
\item \textbf{Chen system:} for $\alpha = \gamma$ with arbitrary real values of the other parameters. 
\item \textbf{L\"u system:} arbitrary real values of the parameters.
\item \textbf{Modified L\"u system:} does not satisfy the Jacobi identity, i.e., $\mathcal{N}$ is not Poisson.
\item \textbf{Qi system:} does not satisfy the Jacobi identity, i.e., $\mathcal{N}$ is not Poisson. 
\end{enumerate}
\end{prop}

\textit{Proof --} By direct calculation. \\

Referring to the discussion in Sec. (\ref{3drev}), given a resistive-Hamiltonian system with $\mathcal{N} = \mathcal{J}\mathcal{R} + \mathcal{R} \mathcal{J}$ being a Poisson matrix, one can always find a function\footnote{In the context of $\mathcal{N}$, we will denote the Hamiltonians with the letter `G' in order to avoid confusion with the Hamiltonians considered earlier in the context of $\mathcal{J}$.} $G \in C^\infty(\mathcal{M}, \mathbb{R})$ and an integrating factor $M$ that allows us to write
\begin{equation}\label{exactness}
\mathbf{N} = \frac{1}{M} (\nabla \overline{G}). 
\end{equation}
As a result, for a chosen $G \in C^\infty(\mathcal{M},\mathbb{R})$, one can get a bi-Hamiltonian system with dynamics described by the vector equation
\begin{equation}\label{biHam0}
M \dot{\mathbf{x}} = (\nabla \overline{G}) \times (\nabla G). 
\end{equation}
To reiterate, starting with a non-conservative system in three dimensions with the resistive-Hamiltonian structure, one can generate another system which is bi-Hamiltonian and conservative provided that $\mathcal{N} = \mathcal{J}\mathcal{R} + \mathcal{R} \mathcal{J}$ is a Poisson matrix. From the vector equation (\ref{biHam0}), it is obvious that a Nambu-Poisson description follows as for an arbitrary function $F \in C^\infty(\mathcal{M},\mathbb{R})$, one can write $\dot{F} = M^{-1} (\nabla F)\cdot [(\nabla \overline{G}) \times (\nabla G)] \equiv \{F, \overline{G}, G\}$ by employing a Nambu three-bracket \cite{nambu0,nambu,nambu2}. Alternatively, one can begin with a conservative Hamiltonian system with Poisson matrix $\mathcal{N}$ and construct from it resistive-Hamiltonian systems by expressing the Poisson matrix as $\mathcal{N} = \mathcal{J} \mathcal{R} + \mathcal{R} \mathcal{J}$; this is discussed in Appendix (\ref{app}). 

\subsection{Two examples}
For the sake of illustration, let us discuss two explicit examples. 

\subsubsection{Reduced three-wave-interaction model}\label{redNex}
We will first consider the example of the reduced three-wave-interaction model which will help clarify our construction. Analogous computations can be performed for the Rabinovich, Chen, and L\"u systems, but not for the modified L\"u and Qi systems, the latter two giving $\mathcal{N}$ which does not satisfy the Jacobi identity and hence does not define a Poisson structure. Now, for the reduced three-wave-interaction model (\ref{red3w}), the components (\ref{Ncomp}) of $\mathbf{N}$ take the following form: 
\begin{equation}\label{red3wNcomponents}
N_x = 0, \quad \quad N_y = 2z^2 - \frac{\gamma z}{2}, \quad \quad N_z = - \gamma y + \frac{\gamma \delta}{2}.
\end{equation}
It is easy to check that $\mathbf{N}\cdot(\nabla \times \mathbf{N}) = 0$ which means one can express $\mathbf{N}$ in the form (\ref{exactness}). Explicitly, one gets
\begin{equation}
\frac{1}{M} \frac{\partial \overline{G}}{\partial y} = 2z^2 - \frac{\gamma z}{2}, \quad \quad \frac{1}{M} \frac{\partial \overline{G}}{\partial z} = - \gamma y + \frac{\gamma \delta}{2},
\end{equation} for functions $\overline{G}$ and $M$ that need to be determined. Eliminating $M$ above, one gets the following partial differential equation:
\begin{equation}
\bigg( \frac{\gamma \delta}{2} - \gamma y \bigg)\frac{\partial \overline{G}}{\partial y}  + \bigg( \frac{\gamma z}{2} - 2z^2 \bigg) \frac{\partial \overline{G}}{\partial z}  = 0.
\end{equation}
One can perform a separation of variables which gives
\begin{equation}
\overline{G} = \left(2y-\delta\right)^{-\lambda/\gamma} \left(1 - \frac{\gamma}{4z} \right)^{2\lambda/\gamma}, \quad \quad M = \frac{\gamma}{4\lambda} \frac{(\gamma z - 4z^2 ) (2y - \delta)}{\overline{G}},
\end{equation} where $\lambda$ is a separation constant which does not affect the dynamical equations. It is straightforward to check that putting the above-mentioned expressions into (\ref{exactness}), one immediately obtains $\mathbf{N}$ whose components are given in (\ref{red3wNcomponents}). With these results, one can define a family of bi-Hamiltonian systems (\ref{biHam0}) for appropriate choices of $G$. For instance, choosing $G = x^2 + y^2 + z$, one obtains the following dynamical equations:
\begin{equation}\label{newbiHamexample}
\dot{x} = 2 \gamma y^2 - \gamma \delta y + 2z^2 - \frac{\gamma z}{2}, \quad \quad \dot{y} = - 2\gamma xy + \gamma \delta x, \quad \quad \dot{z} = - 4 x z^2 +\gamma xz. 
\end{equation}
One thus has a two-parameter bi-Hamiltonian system in which $\dot{G} = 0 = \dot{\overline{G}}$, implying conservation of the two Hamiltonians as expected. 

\subsubsection{L\"u system}
For the L\"u system (\ref{lu}), the components (\ref{Ncomp}) of $\mathbf{N}$ take the following form: 
\begin{equation}\label{LuNcomponents}
N_x = 0, \quad \quad N_y = - \alpha y + \frac{(\beta + \gamma)yz}{\alpha}, \quad \quad N_z = -\alpha z + \frac{\gamma y^2}{\alpha}.
\end{equation}
Although the Poisson vector $\mathbf{N} = (N_x, N_y, N_z)$ satisfies the condition $\nabla \cdot (\nabla \times \mathbf{N}) = 0$ for arbitrary values of the underlying parameters, let us pick $\beta = \gamma$ for the sake of simplicity as this gives $\nabla \times \mathbf{N} = 0$ and consequently from (\ref{exactness}) we get $M = 1$ and 
\begin{equation}
\overline{G} = -\frac{\alpha}{2} (y^2 + z^2) + \frac{\beta y^2 z}{\alpha} . 
\end{equation}
With these results, one can straightforwardly define a family of bi-Hamiltonian systems (\ref{biHam0}) for appropriate choices of $G$. For instance, choosing $G = \frac{x^2}{2} -\alpha z$, one obtains the following dynamical equations:
\begin{equation}\label{newbiHamexample000}
\dot{x} = \alpha^2 y - 2\beta yz, \quad \quad \dot{y} = -\alpha xz + \frac{\beta xy^2}{\alpha}, \quad \quad \dot{z} = \alpha xy - \frac{2 \beta xyz}{\alpha} . 
\end{equation}
One thus has a two-parameter bi-Hamiltonian system in which $\dot{G} = 0 = \dot{\overline{G}}$, implying conservation of the two Hamiltonians as expected. 

\subsection{Jordan-like structures}
To this end, let us notice that $\mathcal{N} = \mathcal{J} \mathcal{R} + \mathcal{R} \mathcal{J}$ resembles the Jordan product \cite{jacobson,jordan}, related closely with the Jordan identity, i.e., $(ab)(aa) = a(b(aa))$ for two elements $a$ and $b$ of a Jordan algebra. However, because $\mathcal{J}$ and $\mathcal{R}$ do not commute, they will generally not be the elements of a Jordan algebra. Nevertheless, one has the following result: 

\begin{prop}
The pair $(\mathcal{J}, \mathcal{R})$ with $\mathcal{J}$ being a Poisson matrix and $\mathcal{R}$ being a resistance matrix satisfies the Jordan-like identity, i.e., 
\begin{equation}\label{jordanid}
(\mathcal{J}\mathcal{R})(\mathcal{J}\mathcal{J}) = \mathcal{J}(\mathcal{R}(\mathcal{J}\mathcal{J})), \quad \quad (\mathcal{R}\mathcal{J})(\mathcal{R}\mathcal{R}) = \mathcal{R}(\mathcal{J}(\mathcal{R}\mathcal{R})). 
\end{equation}
\end{prop}

\textit{Proof --} By direct calculation using the general expressions (\ref{genJR}). \\

This opens up certain possibilities; most importantly that of generating new bi-Hamiltonian systems guided by the structure of Jordan homomorphisms\footnote{Recall that a map between two rings $\phi:R \rightarrow R'$ is a Jordan homomorphism if (a) $\phi(a+b) = \phi(a) + \phi(b)$, (b) $\phi(ab + ba) = \phi(a) \phi(b) + \phi(b)\phi(a)$, for all $a,b \in R$. In the present case, although we do not get a Jordan algebra, we can make use of transformations inspired from Jordan homomorphisms which we will call Jordan-like transformations.}. Let us consider an invertible matrix $\mathcal{T}$ under which $\mathcal{J}$ and $\mathcal{R}$ transform as
\begin{equation}
\mathcal{J}' = \mathcal{T} \mathcal{J} \mathcal{T}^{-1}, \quad \quad \mathcal{R}' = \mathcal{T} \mathcal{R} \mathcal{T}^{-1}. 
\end{equation}
This represents a Jordan-like transformation because 
\begin{equation}
\mathcal{T} (\mathcal{J} \mathcal{R} + \mathcal{R} \mathcal{J}) \mathcal{T}^{-1} = (\mathcal{T} \mathcal{J} \mathcal{T}^{-1}) (\mathcal{T} \mathcal{R} \mathcal{T}^{-1}) + (\mathcal{T} \mathcal{R} \mathcal{T}^{-1}) (\mathcal{T} \mathcal{J} \mathcal{T}^{-1}). 
\end{equation}
Thus, assuming the form of $\mathcal{R}$ to be such that $\mathcal{N} =  \mathcal{J} \mathcal{R} + \mathcal{R} \mathcal{J}$ is a Poisson matrix, one can get another Poisson matrix $\mathcal{N}' =  \mathcal{J}' \mathcal{R}' + \mathcal{R}' \mathcal{J}'$ via a Jordan-like transformation. Of course, $\mathcal{N}'$ has to satisfy the Jacobi identity. In fact, the requirement that $\mathcal{N}'$ is skew-symmetric imposes the restriction that $\mathcal{T}$ must be an orthogonal matrix. As an elementary example, let us consider the bi-Hamiltonian system found from the reduced three-wave-interaction model in Sec. (\ref{redNex}). For a real constant $\Delta \in [-1,1]$, let us define an orthogonal matrix as
\begin{equation}
\mathcal{T} = \begin{pmatrix}
1 & 0 & 0 \\
0 & \Delta & -\sqrt{1-\Delta^2} \\
0 & \sqrt{1-\Delta^2} & \Delta
\end{pmatrix}  \implies 
\mathcal{T}^T \equiv \mathcal{T}^{-1} =  \begin{pmatrix}
1 & 0 & 0 \\
0 & \Delta & \sqrt{1-\Delta^2} \\
0 & - \sqrt{1-\Delta^2} & \Delta
\end{pmatrix}. 
\end{equation}
We thus get from $\mathcal{T} \mathcal{N} \mathcal{T}^{-1} = \mathcal{N}'$, the following: 
\begin{eqnarray}
\mathcal{N}'= \begin{pmatrix}
0 &  -N_z' & N_y'  \\
N_z' & 0 & -N_x' \\
-N_y' & -N_x' & 0
\end{pmatrix}, \end{eqnarray} where
\begin{eqnarray}
\quad \quad 
N_x' &=& 0,  \\
N_y' &=& \bigg( 2z^2 - \frac{\gamma z}{2} \bigg) \Delta - \bigg( - \gamma y + \frac{\gamma \delta}{2} \bigg) \sqrt{1 - \Delta^2}, \nonumber \\
N_z' &=& \bigg( 2z^2 - \frac{\gamma z}{2} \bigg) \sqrt{1 - \Delta^2} + \bigg( - \gamma y + \frac{\gamma \delta}{2} \bigg) \Delta.  \nonumber 
\end{eqnarray}
It is easy to verify the Jacobi identity. Finally taking $G = x^2 + y^2 + z$, one gets the following system as $\dot{\mathbf{x}} = \mathcal{N}' (\nabla G)$:
\begin{eqnarray}
\dot{x} &=& \bigg( 2 \gamma y^2 - \gamma \delta y + 2z^2 - \frac{\gamma z}{2} \bigg) \Delta -  \bigg( 4yz - \gamma yz - \gamma y + \frac{\gamma \delta}{2} \bigg) \sqrt{1 - \Delta^2},  \\
\dot{y} &=& ( - 2 \gamma xy + \gamma \delta x ) \Delta + ( 4 xz^2 - \gamma xz ) \sqrt{1 - \Delta^2}, \nonumber \\
\dot{z} &=& (-4 xz^2 + \gamma xz) \Delta + ( - 2\gamma xy  + \gamma \delta x ) \sqrt{1 - \Delta^2}. \nonumber
\end{eqnarray}
Note that upon putting $\Delta = 1$, one recovers the bi-Hamiltonian system (\ref{newbiHamexample}). The system admits a non-trivial Jacobi last multiplier which may be evaluated straightforwardly and this is therefore not done here. In summary, Jordan-like transformations that map $\mathcal{N} = \mathcal{J} \mathcal{R} + \mathcal{R} \mathcal{J}$ to another Poisson matrix $\mathcal{N}'$ can be used to generate new bi-Hamiltonian systems as illustrated by the simple example above.

\section{Conformal Hamiltonian dynamics on Poisson manifolds}\label{confsec}
Let us now describe conformal Hamiltonian dynamics on Poisson manifolds generalizing the ideas from symplectic geometry\footnote{This generalizes the notion of conformal Hamiltonian dynamics on symplectic manifolds \cite{conf1}, the latter representing infinitesimal non-strictly canonical transformations \cite{conf2}.} \cite{conf1,conf2} (see \cite{ogul1} for discussion in the context of Nambu-Poisson geometry). Consider a Poisson manifold $(\mathcal{M},\Lambda)$ with a Poisson matrix $\mathcal{J}$. A Hamiltonian vector field is thus given by $X_H = \{\cdot, H\}$, where $\{f,g\} = (\nabla f)^T \mathcal{J} (\nabla g)$, as discussed earlier in Sec. (\ref{revsec}). Consider now the dilatation vector field $\Gamma$ which in local coordinates reads
\begin{equation}
\Gamma = x_i \frac{\partial}{\partial x_i}, 
\end{equation} i.e., $\Gamma$ acting on a homogeneous polynomial returns the degree of homogeneity (say, $m$) as $\Gamma (p(x_1,x_2,\cdots,x_n)) = m p(x_1,x_2,\cdots,x_n)$. In other words, $\Gamma$ is just the Euler operator. 

\begin{axiom}\label{defconf}
Consider a Poisson manifold $(\mathcal{M}, \Lambda)$ with the Poisson matrix $\mathcal{J}$ which describes a volume-preserving Hamiltonian dynamics as $X_H = \{\cdot, H\}$, i.e., ${\rm div}(X_H) = 0$. One can define a conformal Hamiltonian vector field as
\begin{equation}\label{confvec}
X^a_H = \Lambda(dH, \cdot) + a \Gamma, 
\end{equation} where $H \in C^\infty(\mathcal{M}, \mathbb{R})$, $a \in \mathbb{R}$, and $\Gamma$ is the dilatation/Euler operator. 
\end{axiom}

That is to say, one can write $X^a_H = X_H + a \Gamma = \{\cdot, H\} + a \Gamma$. The corresponding dynamics has a non-trivial divergence; while the Poisson part of the evolution contributes zero divergence (as assumed above), the dilatation vector field provides a non-trivial contribution so that one has ${\rm div} (X^a_H) = n a$. Thus, the following is true:

\begin{prop}
A conformal Hamiltonian vector field $X^a_H$ for $H \in C^\infty(\mathcal{M}, \mathbb{R})$ and $a \in \mathbb{R}$ satisfies
\begin{equation}
\pounds_{X^a_H} \Omega = (n a) \Omega, 
\end{equation} where $\Omega = dx_1 \wedge dx_2 \wedge \cdots \wedge dx_n$ is a volume-form on $\mathcal{M}$. 
\end{prop}

Note that upon putting $a = 0$ one recovers the standard Hamiltonian vector field. The Hamiltonian is not conserved under the flow of $X^a_H$ for $a \neq 0$ as $X^a_H (H) = a \Gamma(H)$. Thus, a conformal Hamiltonian vector field has the typical characteristics required for describing a non-conservative time evolution. A key feature of a conformal Hamiltonian vector field is the constancy of its divergence, i.e., although the divergence is non-zero, it is a constant. In what follows, we will demonstrate how some of our previously-considered examples can be described alternatively using conformal Hamiltonian dynamics on Poisson manifolds, i.e., without resorting to resistance matrices. 

\subsection{Reduced three-wave-interaction model}
For the reduced three-wave-interaction model given by the system (\ref{red3w}), the dynamics can be described by conformal Hamiltonian dynamics if one chooses $\gamma = -2$ in which case the Hamiltonian and the Poisson matrix are given by
\begin{equation}
H = x^2 + y^2 + z, \quad \quad 
\mathcal{J} = \begin{pmatrix}
0 &  -y + \frac{\delta}{2} &  z \\
y - \frac{\delta}{2} &  0 &  0 \\
-z & 0 & 0 \\
\end{pmatrix}.
\end{equation}
The phase trajectories of the system (\ref{red3w}) are then the integral curves of the conformal Hamiltonian vector field $X^{-2}_H$, i.e., with $a = -2$ in (\ref{confvec}).

\subsection{Chen system}
The Chen system (\ref{chen}) can be described by conformal Hamiltonian dynamics for $\alpha = \beta = -\gamma$ in which case the Hamiltonian and the Poisson matrix are given by\begin{equation}
H = \frac{x^2}{2} - \alpha z, \quad \quad 
\mathcal{J} = \begin{pmatrix}
0 &  -\gamma + z &  -y \\
\gamma -z &  0 &  x \\
y  & -x & 0 \\
\end{pmatrix}. \end{equation}
The phase trajectories of the system (\ref{chen}) are then the integral curves of the conformal Hamiltonian vector field $X^{-\alpha}_H$, i.e., with $a = -\alpha$ in (\ref{confvec}).

\subsection{L\"u system}
Considering the L\"u system (\ref{lu}), a description in terms of conformal Hamiltonian dynamics is possible for $\alpha = \beta = -\gamma$ in which case the Hamiltonian and the Poisson matrix are given by\begin{equation}
H = \frac{x^2}{2} - \alpha z, \quad \quad 
\mathcal{J} = \begin{pmatrix}
0 &   z &  -y \\
 -z &  0 &  0 \\
y  & 0 & 0 \\
\end{pmatrix}. \end{equation}
The phase trajectories of the system (\ref{lu}) are then the integral curves of the conformal Hamiltonian vector field $X^{-\alpha}_H$, i.e., with $a = -\alpha$ in (\ref{confvec}).

\section{Conclusions}\label{dissec}
In this paper, we discussed the Hamiltonian formulations of some three-dimensional systems following up on the related developments \cite{ogul,ogul1,ogul2,ogul3}. While the focus in previous developments was on bi-Hamiltonian and Nambu-metriplectic descriptions, we introduced in this paper a novel scheme that we dubbed resistive-Hamiltonian systems (inspired from port-Hamiltonian systems) in which the dynamics on a Poisson manifold as described by a Poisson matrix $\mathcal{J}$ is supplemented by a symmetric resistance matrix $\mathcal{R}$ that describes the non-conservative part of the time evolution; in this sense, it has similarities with the metriplectic formalism \cite{morr,morr1,morr2,guha_metri} although the latter requires additional integrability conditions and is distinct from the present approach. We also discussed higher-degree Poisson matrices defined as $\mathcal{J} \mathcal{R} + \mathcal{R} \mathcal{J}$ which may allow one to define new bi-Hamiltonian systems starting with a known non-conservative system described by the framework of resistive-Hamiltonian systems. Finally, we showed that some of the systems, namely, the reduced three-wave-interaction model, the Chen system, and the L\"u system can alternatively be described by conformal Hamiltonian dynamics (by appropriately choosing the underlying parameters) in which the Hamiltonian vector field which is taken to be of the volume-preserving nature is supplemented with the dilatation/Euler vector field with the latter describing the non-conservative part of the dynamics. While our focus in this paper has been on the Hamiltonian aspects of three-dimensional systems, future directions emerging from here would include on the one hand, studying the dynamical properties (including fixed points, their stability, etc.) of the new bi-Hamiltonian systems that can be found from the construction of higher-degree Poisson matrices, while on the other hand, extending the existing framework to describe dynamical systems within the premise of port-Hamiltonian systems with external ports.

\section*{Acknowledgements} A.G. thanks Akash Sinha for discussions and Ranita Mudi for verifying (\ref{jordanid}) using Mathematica software. A.G.C. and P.G. thank O\u{g}ul Esen and Hasan G\"umral for related discussions.

\appendix

\section{Reverse-engineering non-conservative systems}\label{app}
\numberwithin{equation}{section}
If one knows a certain conservative system on a Poisson manifold with a Poisson matrix $\mathcal{N}$ and a Hamiltonian $G$, then one can find appropriate matrices $\mathcal{J}$ and $\mathcal{R}$ with the former being a Poisson matrix such that the two when taken together can define a resistive-Hamiltonian (non-conservative) system. As an example, let us consider the free Euler rotations in three dimensions which was the classic problem taken by Nambu to illustrate his three-bracket construction \cite{nambu0}. For this problem, one has two known conserved quantities, namely, the kinetic energy of the rotor and its total angular momentum:
\begin{equation}
K = \frac{L_x^2}{2I_x} + \frac{L_y^2}{2I_y} +  \frac{L_z^2}{2I_z}, \quad \quad L^2 = L_x^2 + L_y^2 + L_z^2,
\end{equation} where $(L_x,L_y,L_z)$ are the components of the angular momentum and $(I_x,I_y,I_z)$ are the moments of inertia with the coordinate axes being chosen along the principal axes of the rotor so the cross terms of the inertia tensor are trivial. Taking $(L_x,L_y,L_z)$ to be the coordinates on a three-dimensional phase space, it can be endowed with a Poisson vector given by
\begin{equation}
\mathbf{N} = \frac{1}{2} (\nabla \overline{G}) , \quad \quad \overline{G} = L^2,
\end{equation} and which immediately gives $\mathbf{N} = (L_x, L_y, L_z)$, i.e., one has the Poisson matrix
 \begin{equation}
 \mathcal{N} = \begin{pmatrix}
0 & -L_z & L_y \\
L_z & 0 & -L_x \\
-L_y & L_x & 0
\end{pmatrix}. 
 \end{equation}
 Obviously, the Jacobi identity is satisfied. The Euler equations are obtained from $\dot{\mathbf{x}} = \mathcal{N} (\nabla G)$, where $\mathbf{x} = (L_x \quad L_y \quad L_z)^T$ and $G$ is the identified with the total kinetic energy $K$. The result is the familiar system of equations
 \begin{equation}\label{Eulereq}
 \dot{L}_x = \bigg(\frac{1}{I_z} - \frac{1}{I_y} \bigg) L_y L_z, \quad \quad  \dot{L}_y = \bigg(\frac{1}{I_x} - \frac{1}{I_z} \bigg) L_z L_x, \quad \quad  \dot{L}_z = \bigg(\frac{1}{I_y} - \frac{1}{I_x} \bigg) L_x L_y. 
 \end{equation}
 Let us now find a non-conservative system from here. Notice that $\mathcal{N}$ can be expressed in the form $\mathcal{N} = \mathcal{J} \mathcal{R} + \mathcal{R} \mathcal{J}$ such that
 \begin{equation}
  \mathcal{J} = \begin{pmatrix}
0 & -L_z & L_y \\
L_z & 0 & -L_x \\
-L_y & L_x & 0
\end{pmatrix}, \quad \quad  \mathcal{R} = \frac{1}{2} \begin{pmatrix}
1 & 0 & 0 \\
0 & 1 & 0 \\
0 & 0 & 1
\end{pmatrix}. 
 \end{equation}
Notice that $\mathcal{J}$ coincides with $\mathcal{N}$. Equipped now with both a Poisson matrix $\mathcal{J}$ and a resistance matrix $\mathcal{R}$, one can define a resistive-Hamiltonian system as $\dot{\mathbf{x}} = (\mathcal{J} - \mathcal{R}) (\nabla H)$. Taking $H = K$, i.e., the kinetic energy of the rotor, one gets the following system:
 \begin{equation}\label{Eulerdiss}
 \dot{L}_x = - \frac{L_x}{2I_x} + \bigg(\frac{1}{I_z} - \frac{1}{I_y} \bigg) L_y L_z,  \quad  \dot{L}_y = - \frac{L_y}{2I_y} + \bigg(\frac{1}{I_x} - \frac{1}{I_z} \bigg) L_z L_x, \quad  \dot{L}_z = - \frac{L_z}{2I_z} + \bigg(\frac{1}{I_y} - \frac{1}{I_x} \bigg) L_x L_y.
 \end{equation} One can easily check that the kinetic energy is not conserved under the above-mentioned dynamics unlike in the case of the system (\ref{Eulereq}) for which the kinetic energy is conserved. We can call (\ref{Eulerdiss}) a system of `dissipative' Euler equations. It is straightforward to find out many more examples of the above-mentioned construction.


\begin{thebibliography}{99}

\bibitem{arnold}V. I. Arnold, {\it Mathematical Methods of Classical Mechanics}, Graduate Texts in Mathematics Series, vol. 60 (2nd ed.), Springer (1989).

\bibitem{poisson1} I. Vaisman, {\it Lectures on the Geometry of Poisson Manifolds}, Progress in Mathematics Series, vol. 118, Springer (1994). 

\bibitem{morr}P. J. Morrison, Rev. Mod. Phys. \textbf{70}, 467 (1998).

\bibitem{fluid1} A. I. Dyachenko, P. M. Lushnikov, and V. E. Zakharov, J. Fluid Mech. \textbf{869}, 526 (2019).

\bibitem{fluid2} E. Tassi, Physica D \textbf{437}, 133338 (2022). 

\bibitem{morr1}P. J. Morrison, Phys. Lett. A \textbf{100}, 423 (1984).

\bibitem{morr2}P. J. Morrison, Physica D \textbf{18}, 410 (1986).

\bibitem{stat} M. Pavelka, V. Klika, O. Esen, and M. Grmela, Physica D \textbf{335}, 54 (2016). 

\bibitem{elas1} B. J. Edwards and A. N. Beris, J. Phys. A: Math. Gen. \textbf{24}, 2461 (1991). 

\bibitem{elas2} K.-C. Chen, Int. J. Solids Struct. \textbf{44}, 7715 (2007). 

\bibitem{nambu0} Y. Nambu, Phys. Rev. D \textbf{7}, 2405 (1973).

\bibitem{nambu} L. Takhtajan, Commun. Math. Phys. \textbf{160}, 295 (1994).

\bibitem{nambu2} P. Guha, Nonlinear Anal. Theory Methods Appl. \textbf{65}, 2025 (2006).

\bibitem{3w2} A. S. Pikovskii and M. I. Rabinovich, Sov. Sci. Rev. C: Math. Phys. Rev. \textbf{2}, 165 (1981).

\bibitem{3w1} H. J. Giacomini, C. E. Repetto, and O. P. Zandron, J. Phys. A: Math. Gen. \textbf{24}, 4567 (1991).

\bibitem{chen} G. Chen and T. Ueta, Int. J. Bifurc. Chaos \textbf{9}, 1465 (1999). 
 
 \bibitem{lu1} J. L\"u and G. Chen, Int. J. Bifurc. Chaos \textbf{12}, 659 (2002).

\bibitem{lu2} J. L\"u, G. Chen, and S. Zhang, Chaos Solit. Fractals \textbf{14}, 669 (2002).

\bibitem{qi} G. Qi, G. Chen, S. Du, Z. Chen, and Z. Yuan, Physica A \textbf{352}, 295 (2005). 

\bibitem{t} G. Tigan and D. Opri\c{s}, Chaos Solit. Fractals \textbf{36}, 1315 (2008).

\bibitem{gumralimp} H. G\"umral and Y. Nutku, J. Math. Phys. \textbf{34}, 5691 (1993).

\bibitem{gumral} H. G\"umral, arXiv:1003.0343.

\bibitem{gao} P. Gao, Phys. Lett. A \textbf{273}, 85 (2000).

\bibitem{ogul} O. Esen, A. Ghose Choudhury, and P. Guha, Int. J. Bifurc. Chaos \textbf{26}, 1650215 (2016).

\bibitem{ogul1} O. Esen and P. Guha, J. Geom. Phys. \textbf{127}, 32 (2018).

\bibitem{ogul2} O. Esen, A. Ghose Choudhury, and P. Guha, Theor. Appl. Mech. \textbf{44}, 15 (2017).

\bibitem{ogul3} O. Esen and P. Guha, Int. J. Geom. Methods Mod. Phys. \textbf{17}, 2050042 (2020). 

\bibitem{JLM} E. T. Whittaker, {\it A Treatise on the Analytical Dynamics of Particles and Rigid Bodies}, Cambridge University Press (1988).

\bibitem{JLM1} P. Guha and A. Ghose Choudhury, Rev. Math. Phys. \textbf{25}, 1330009 (2013).

\bibitem{port} A. van der Schaft and D. Jeltsema, {\it Port-Hamiltonian Systems Theory: An Introductory Overview}, Foundations and Trends\textregistered ~in Systems and Control Series, vol. 1, Now Publishers (2014).

\bibitem{port1} V. Duindam, A. Macchelli, S. Stramigioli, and H. Bruyninckx (Eds.), {\it Modeling and Control of Complex Physical Systems: The Port-Hamiltonian Approach}, Springer (2009). 

\bibitem{jacobson} N. Jacobson, {\it Structure and Representations of Jordan Algebras}, AMS Colloquium Publications, vol. 39, American Mathematical Society (1968). 

\bibitem{jordan} T. A. Springer, {\it Jordan Algebras and Algebraic Groups}, Classics in Mathematics Series, Springer (1998). 

\bibitem{conf1} R. McLachlan and M. Perlmutter, J. Geom. Phys. \textbf{39}, 276 (2001). 

\bibitem{conf2} J. F. Cari\~nena, F. Falceto, and M. F. Ra\~nada, J. Geom. Mech. \textbf{5}, 151 (2013). 

\bibitem{AG} A. Ghosh, Phys. Scr. \textbf{100}, 035220 (2025). 

\bibitem{guha_metri} P. Guha, J. Math. Anal. Appl. \textbf{326}, 121 (2007).

\end{thebibliography}
\end{document}